\documentclass[10pt,conference]{IEEEtran}
\IEEEoverridecommandlockouts
\usepackage{cite}
\usepackage{amsmath,amssymb,amsfonts}
\usepackage{algorithmic}
\usepackage{graphicx}
\usepackage{textcomp}
\usepackage{xcolor}
\usepackage{balance}

\usepackage{hyperref}
\usepackage{todonotes}
\usepackage{booktabs}
\usepackage{siunitx}
\usepackage{multirow}
\usepackage{subcaption}
\usepackage{tcolorbox}
\def\BibTeX{{\rm B\kern-.05em{\sc i\kern-.025em b}\kern-.08em
		T\kern-.1667em\lower.7ex\hbox{E}\kern-.125emX}}
\begin{document}
	
	\title{Collaborative, Code-Proximal\\ Dynamic Software Visualization\\ within Code Editors}

	\author{\IEEEauthorblockN{Alexander Krause-Glau}
		\IEEEauthorblockA{\textit{Software Engineering Group} \\
			\textit{Kiel University}\\
			Kiel, Germany \\
			akr@informatik.uni-kiel.de}
		\and
		\IEEEauthorblockN{Wilhelm Hasselbring}
		\IEEEauthorblockA{\textit{Software Engineering Group} \\
			\textit{Kiel University}\\
			Kiel, Germany \\
			wha@informatik.uni-kiel.de}
	}
	
	\maketitle
	\thispagestyle{plain}
	
	\begin{abstract}
		Software visualizations are usually realized as standalone and isolated tools that use embedded code viewers within the visualization.
		In the context of program comprehension, only few approaches integrate visualizations into code editors, such as integrated development environments.		
		This is surprising since professional developers consider reading source code as one of the most important ways to understand software, therefore spend a lot of time with code editors.	
		
		In this paper, we introduce the design and proof-of-concept implementation for a software visualization approach that can be embedded into code editors.
		Our contribution differs from related work in that we use dynamic analysis of a software system's runtime behavior.
		Additionally, we incorporate distributed tracing.
		This enables developers to understand how, for example, the currently handled source code behaves as a fully deployed, distributed software system.
		Our visualization approach enhances common remote pair programming tools and is collaboratively usable by employing shared code cities.
		As a result, user interactions are synchronized between code editor and visualization, as well as broadcasted to collaborators.
		
		To the best of our knowledge, this is the first approach that combines code editors with collaboratively usable code cities.
		Therefore, we conducted a user study to collect first-time feedback regarding the perceived usefulness and perceived usability of our approach.
		We additionally collected logging information to provide more data regarding time spent in code cities that are embedded in code editors.
		Seven teams with two students each participated in that study.		
		The results show that the majority of participants find our approach useful and would employ it for their own use.
		We provide each participant’s video recording, raw results, and all steps to reproduce our experiment as supplementary package.
		Furthermore, a live demo of our tool is available online.\footnote{\url{https://code.explorviz.dev}}
		We invite other researchers to extend our open-source software.\footnote{\url{https://github.com/ExplorViz}}
		Video URL: \url{https://youtu.be/3qZVSehnEug}
	\end{abstract}
	
	
	\section{Introduction}\label{sec:introduction}
Source code comprehension is still the primary method to come to an understanding of a software system's behavior~\cite{xia2018}.
This is not unexpected, because developers are trained to recognize recurring patterns and resulting behavior in source code.
They even might spend most of their development time in integrated development environments (IDE)~\cite{tiarks2011}.
However, navigation in IDEs leads to a redundant but unavoidable overhead~\cite{ko2006} and in terms of software visualization (SV) developers are concerned about the context switch caused by standalone SV tools~\cite{sensalire2007}.
As a result, code proximity is a necessary property for SV~\cite{bassil2001,kienle2007} to succeed in its intended area, i.e., professional software development.
Code proximity means the ability of the visualization tool to provide easy and fast access to the original, underlying source code~\cite{Lanza2003}.
In this context, research approaches have been shown in the past that embed SV into code editors and IDEs (both from now on referred to as \emph{code editor}) to link source code with its visualization~\cite{lintern2003,balogh2015codemetropolis,sulir2018,kurbatova2021}.

In this paper, we introduce our collaboratively usable SV approach that can be embedded in code editors.
In comparison to related approaches, we use dynamic analysis as source for rendering three-dimensional code cities~\cite{knight1999City, wettel2007}.
The SV is linked directly to the source code that is under development within the code editor and vice versa.
Therefore, we directly connect runtime behavior with the related program elements, for example, Java methods.
User interactions are synchronized between code editor and visualization, as well as broadcasted to collaborators.
As proof of concept, we implemented a Visual Studio Code\footnote{\url{https://code.visualstudio.com}} (VS Code) extension that realizes our design.
We conducted a first-time user study to collect feedback regarding the perceived usefulness and perceived usability of our approach.
Furthermore, we collected logging information to provide more data regarding usage statistics of SV that are embedded into code editors.
In this study, seven teams with  two students each collaboratively used our approach in an onboarding-related scenario.
Overall, the results show a highly rated usefulness.

\begin{figure*}
	\centering
	\includegraphics[width=\textwidth]{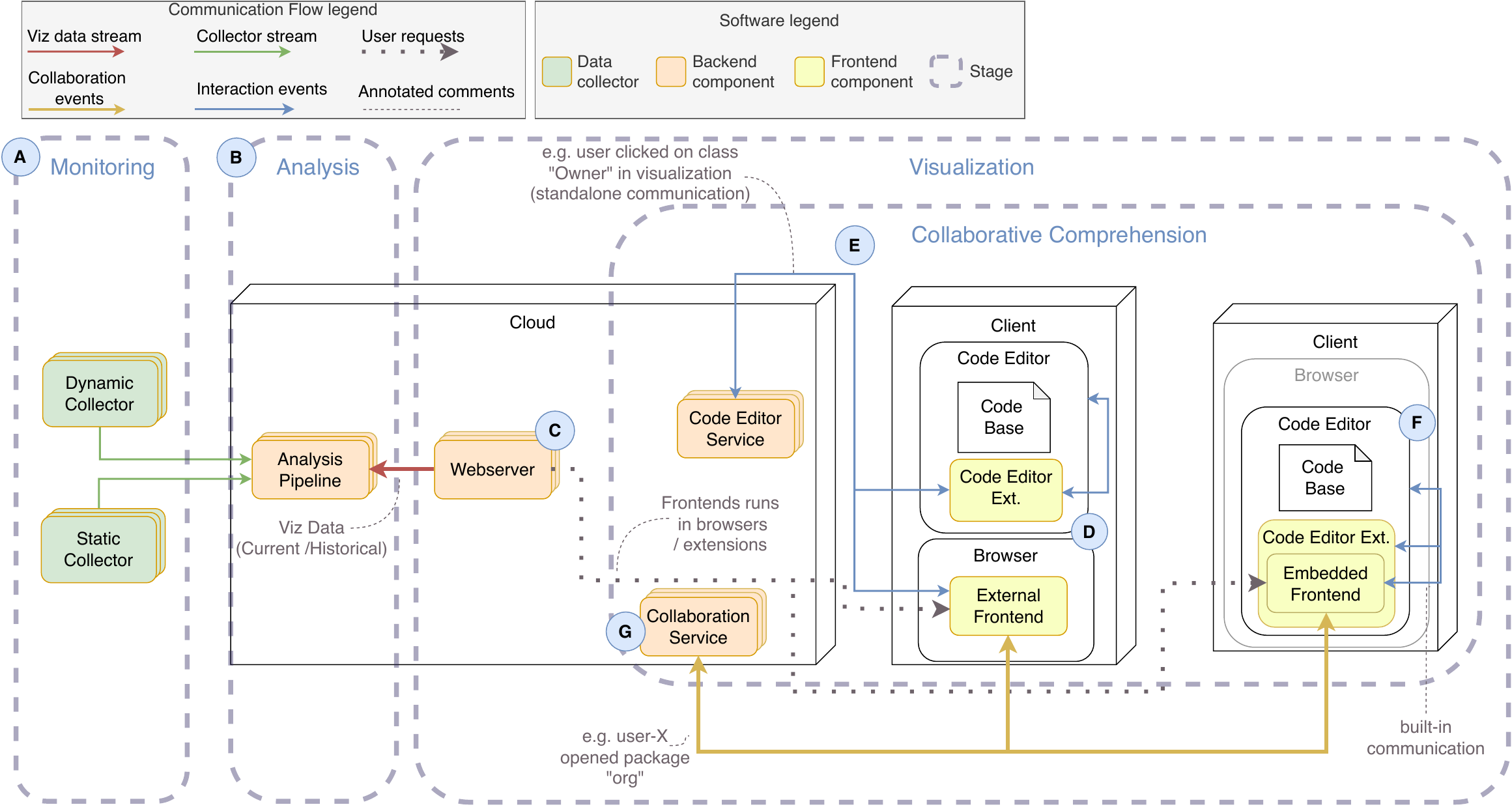}
	\caption{(Simplified) Architectural design of our approach.}
	\label{fig:conceptual-design}
\end{figure*}

The remainder of this paper is structured as follows. 
Section~\ref{sec:approach} presents the architectural overview and proof of concept implementation for our approach.
We proceed by introducing the envisioned usage scenarios for our approach in Section~\ref{sec:scenarios}.
Afterwards, Section~\ref{sec:evaluation} explains our experimental setup.
Section \ref{sec:results} presents and discusses the results of our study.
Then, Section~\ref{sec:related-work} introduces related work.
Finally, we conclude this paper and present future work in Section~\ref{sec:conclusions}.

	\begin{figure*}
	\includegraphics[width=\textwidth]{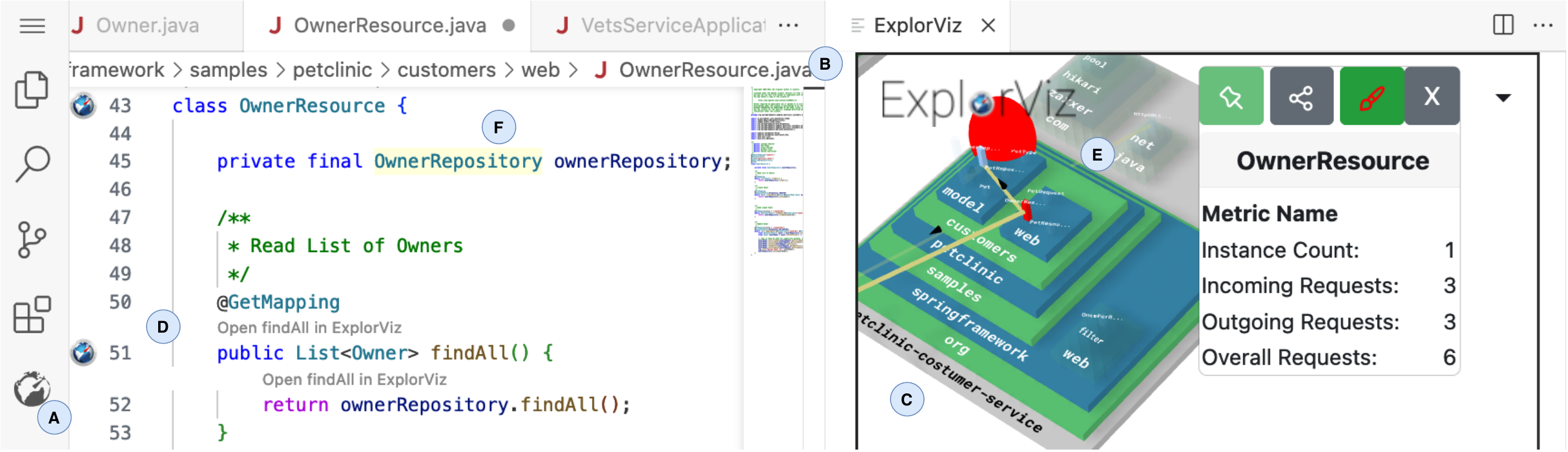}
	\caption{Proof of concept implementation -- The editor of VS Code displays a Java class. The ExplorViz extension visualizes the associated runtime behavior and adds visual functions to the editor to directly link source code and visualization. }
	\label{fig:implementation}
\end{figure*}

\section{Approach}\label{sec:approach}
In this section, we present the architectural design and proof of concept implementation for this research work.
For that, we build upon our previously published approach named \textit{Software Visualization as a Service} (SVaaS), i.e., providing an online-accessible and on-demand service for collaborative program comprehension using SV.
Due to space constraints, we refer readers to~\cite{krauseglau2022ic2e} for a description of the basic concepts of our approach.

\subsection{Architectural Design}\label{sec:design}
Figure~\ref{fig:conceptual-design} shows (a simplified overview of) our approach's architectural design.
It is technology independent with the exception of a browser-based SV component.
As shown, it is divided into four stages (blue-striped areas).
Figure~\ref{fig:conceptual-design}-A and Figure~\ref{fig:conceptual-design}-B depict the monitoring and analysis stages, respectively.
These are the foundation of our SVaaS concept.
The analysis pipeline for example can be horizontally scaled out to handle varying load of concurrent users, therefore positively influence the effectiveness of the overall tool~\cite{sensalire2008}.
Although data acquisition, analysis, and cloud technologies are important aspects of our concept, a detailed explanation is beyond the scope of this paper.
Therefore, we refer readers to~\cite{krauseglau2022ic2e} for details and focus on the remaining two stages.

The \textsf{Webserver} (Figure~\ref{fig:conceptual-design}-C) serves the static files that comprise the web-based SV, i.e., CSS, JavaScript, and HTML.
Furthermore, it acts as reverse proxy for clients to connect to the backend services, e.g., to obtain data to be visualized.
Users now have two options to link the SV with their code editor:
\begin{itemize}
	\item For the first option, they can use the standalone SV that runs inside of their web browser and connects to an extension in their code editor (Figure~\ref{fig:conceptual-design}-D).
The latter acts as gateway between code editor and SV.
This is similar to the `classic' approach for code viewers embedded into SVs and relates to many other works (see Section~\ref{sec:related-work}).
Interactions that should be linked between code editor and SV, e.g., `open a class file in the code editor when the related visualization entity was clicked', are synchronized by the \textsf{Code  Editor Service}  (Figure~\ref{fig:conceptual-design}-E).
	\item For the second option, users can install an extension in their code editor that already includes the \textsf{Frontend} (Figure~\ref{fig:conceptual-design}-F).
In this case, we do not need an external service to synchronize interaction events, but use a built-in communication mechanism between code editor and its extension.
Therefore, we reduce the context switch overhead that occurs when switching from SV to code editor and vice versa~\cite{sensalire2007}.
Another advantage of the second option is that it can also be installed in cloud-based code editors that run in browsers.
This can be beneficial in some use cases, e.g., onboarding of new developers, as shown in Section~\ref{sec:scenarios}.
\end{itemize}
\newpage
Regardless of the selected option, users can collaboratively use the SV.
To achieve this, the  \textsf{Collaboration Service} (Figure~\ref{fig:conceptual-design}-G) broadcasts events, e.g., `user X opened package Y', to all clients of the same session except the one that triggered an event~\cite{krauseglau2022vissoft}.
The clients then apply the received events to their SV, therefore synchronize their states.

\subsection{Proof of Concept Implementation}\label{sec:implementation}
We have prototyped our approach within our SV tool ExplorViz.\footnote{\url{https://explorviz.dev}}
Our tool's development commenced in 2012~\cite{hasselbring2020} and focused on several aspects throughout time, such as development concerns~\cite{zirkelbach2019,krause2020} and extended reality~\cite{Fittkau2015,krauseglau2022vissoft}.
More recently, we use ExplorViz to research collaborative software visualization in the context of program comprehension~\cite{krauseglau2022vissoft, krauseglau2022ist}.
ExplorViz currently uses dynamic analysis as source for the visualization.
Our depicted SV is configured to visualize the aggregated runtime behavior of ten seconds~\cite{krauseglau2022ic2e}.

Figure~\ref{fig:implementation} shows a screenshot of our prototype implementation.
We developed a VS Code extension that realizes the previously mentioned design.
It can be used as gateway to link the external SV to the code editor or provide the embedded SV instead.
Due to space constraints, we focus on the latter and refer readers to our supplementary video.
The extension uses an HTML iFrame to embed the web-based \textsf{Frontend}, therefore SV, in VS Code (see Figure~\ref{fig:conceptual-design}-F on the previous page).
The embedded SV can be switched on or off via the ExplorViz logo button (Figure~\ref{fig:implementation}-A).
It is automatically placed in a new editor group next to the source code (Figure~\ref{fig:implementation}-B).
Users can select one of their (currently or previously) analyzed software systems (as shown in the supplementary video) and open the related SV.
The latter is provided as three-dimensional code cities using Three.js\footnote{\url{https://threejs.org}} for rendering (Figure~\ref{fig:implementation}-C).
The embedded \textsf{Frontend} uses cross-origin communication based on the JavaScript Window object to interact with VS Code.
Therefore, we do not need an external service that synchronizes the interaction events as it is the case when using the external \textsf{Frontend} or as shown in related works (see Section~\ref{sec:related-work}).
Every tenth second the \textsf{Frontend} triggers a SV update.
For that, it obtains the latest runtime data for the selected software system from the analysis pipeline and updates the visualization if required.
Furthermore, the \textsf{Frontend} sends new data to VS Code, which then highlights Java classes and methods that have been used in the aggregated runtime behavior.
This is shown by the gutter icons and code lenses in Figure~\ref{fig:implementation}-D.
Users can click on a code lens to focus the related entity in the SV, e.g., a high-rise building visualizing a Java class.
Vice versa, pressing for example on a communication line will cause the file to open and focus on the related method in VS Code.
In terms of working together, users can join or host a collaborative session from within the embedded \textsf{Frontend} and use the collaborative features of the SV, e.g., pinging or shared popups (Figure~\ref{fig:implementation}-E), to interact with each other (please see~\cite{krauseglau2022vissoft} for more details).
Furthermore, a collaborative session also enables remote pair programming.
For VS Code in general, developers can for example use Microsoft's LiveShare extension for VS Code.
LiveShare has great features and usability, but uses Microsoft servers that might be not available in the future or cannot be used due to compliance concerns.
For the sake of our evaluation's reproducibility, we therefore decided against using an available product such as Microsoft's LiveShare, but developed our own solution (for the user study).
This can be seen in Figure~\ref{fig:implementation}-F where the live text selection of another user is depicted (as yellow background of \texttt{OwnerRepository}).
These text selection events are synchronized by an implementation of the external \textsf{Code Editor Service} (Figure~\ref{fig:conceptual-design}-E) using WebSockets for almost real-time communication.
	\section{Envisioned Usage Scenarios}\label{sec:scenarios}
Besides using advanced (web) technologies, our approach can be differentiated from related work by the use of dynamic analysis and collaborative SV features.
Therefore, we now introduce envisioned usage scenarios that may follow from our approach and related future works.

\subsection*{Scenario 1 (SC1): Facilitate the Onboarding Process}
In professional software development, companies utilize different techniques for the onboarding process of new developers.
Peer support, product overview, and simple tasks are perceived as useful in that context~\cite{buchan2019}, while finding documentation and technical issues, e.g., setting up a development environment, impede the onboarding process, especially for remote work~\cite{rodeghero2021}.
We envision a scenario where cloud-based code editors with embedded SVs are prepared to guide new developers step-by-step through a software system's behavior.
Users click on a use case of the analyzed (distributed) target system and understand its unfolding via SV.
Furthermore, increasingly large portions of the source code (e.g., depending on experience) are directly linked to SV entities.
This allows developers to understand which portion of the source code acts in which use cases.
The approach can then be used for task-oriented onboarding, where developers also face small tasks to comprehend the software~\cite{buchan2019, an2021}.
At any time, users can invite other developers for collaborative comprehension or their mentor and ask for help.
Next to voice communication, participants use collaborative features such as synchronized text selection and shared information popups to interact and exchange~\cite{krauseglau2022vissoft}.
 

\subsection*{Scenario 2 (SC2): Highlight changes during code reviews}
Feature requests and resulting change-based code reviews are commonly used in professional software development~\cite{baum2017}.
However, reviewers tend to give vacuous feedback and generally report on review tools' limitations when used in complex scenarios~\cite{dong2021}.
In this context, we see another potential usage scenario for our approach that we outline in the following.
A team member is supposed to review source code changes of a colleague.
To do this, he or she can click on a link inside of the pull request that opens a prepared, cloud-based code editor with an embedded SV of the new program behavior (due to the source code change). 
Source code changes are color-coded in the IDE.
For understanding the program behavior, it is possible to switch between old and new program behavior in the SV by pressing a button.
The colleague who issued the pull request can be invited to the session such that the changes can also be discussed together.

\subsection*{Scenario 3 (SC3): Integrate Runtime Information into Development Activities}
Staging environments are used to test software systems in a production-like environment.
We envision code editors informing selected developers about performance problems of a software system installed (e.g., in the staging area).
A developer can click on this notification to open the embedded SV. 
The visualization depicts the runtime behavior which includes the performance problem.
It also highlights the entity that introduces the problem, e.g., a method call that took too long to finish.
Based on this, developers get runtime information displayed in their code editor and can analyze affected code lines.


	\section{Experiment Design and Demographics}\label{sec:evaluation}
Effectiveness is one of the most common properties used to evaluate SV approaches.
In that context, Merino et al.~\cite{merino2018SLR} present a systematic literature review of SV evaluation.
Their work analyzes the literature body of full papers that were published in the SOFTVIS/VISSOFT conferences, resulting in the examination of 181 papers.
The authors focus on evaluations that validate the effectiveness of their presented approach.
It is mentioned that multiple evaluations omit other variables that can contribute to or generally influence the effectiveness~\cite{feitelson2022CodeComprehensionThreats}, such as recollection and emotions.
We share this opinion and argue that we must first evaluate properties such as perceived usefulness, perceived usability, or feature requests to potentially refine a new, exploratory approach.
Only afterwards, we should evaluate effectiveness and efficiency with a sufficiently large number of participants in controlled experiments~\cite{gall1996}.
As a result, we decided to conduct an exploratory user-study first.
We designed an experiment in which participants use and evaluate our approach in a task-oriented onboarding process, i.e., in a scenario similar to SC1 (see Section~\ref{sec:scenarios}).
In the future, we will also evaluate our approach in other scenarios by using a similar experiment.
In this paper however, we developed the experiment with a focus on SC1 due to the approach's prototype implementation, the exploratory nature of the study, and the duration of a single experiment run.
As a result, our research questions (RQ) are not concerned about effectiveness or efficiency.
Instead, we focus on several aspects to gather qualitative feedback and quantitative results, such as time spent in the embedded SV, to gain first insights into the use of our approach:

\begin{itemize}
	\item \textbf{RQ1}: How do subjects use the embedded SV and code editor during task solving?
	\item \textbf{RQ2}: Is the code editor perceived as more useful than the embedded SV?
	\item \textbf{RQ3}: Do subjects recognize the usefulness of collaborative SV features for specific tasks?
	\item \textbf{RQ4}: What is the general perception of the usefulness and usability of the approach?
	\item \textbf{RQ5}: Is the approach perceived as useful in the envisioned usage scenarios?
\end{itemize}
We again emphasize that the findings of this contribution should be seen as first insights and indicators for refinements rather than statistically grounded results.
However, by answering the research question, we can derive the following main \textbf{contributions} of our evaluation:

\begin{itemize}
	\item Further insights regarding the perceived usefulness of software cities to comprehend runtime behavior.
	\item First quantitative and qualitative results regarding the perceived usefulness, perceived usability, and usage time for collaborative, code-proximal software cities.
	\item A supplementary package containing the evaluation's raw results, screen recordings of all participants, and detailed instructions as well as software packages for reproduction~\cite{krause2023CollabSuppl}.
\end{itemize}
In the following, we now present the participants' demography and our experiment's procedure.
\begin{figure}[t]
	\includegraphics[width=\linewidth]{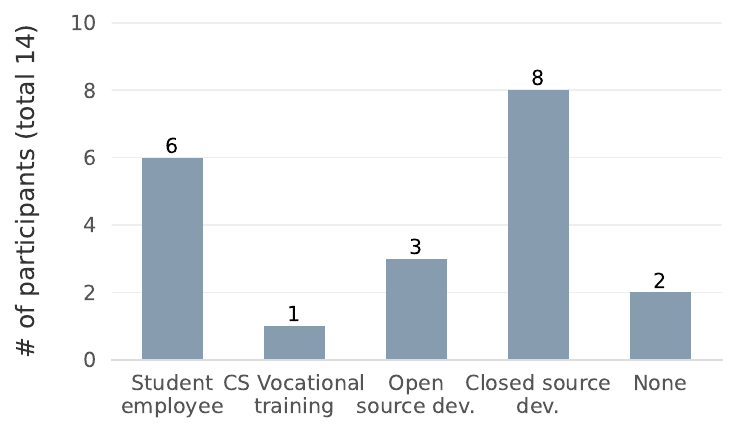}
	\caption{Participants' reported experiences with software development based on work experiences (multi-choice).}
	\label{fig:se-experiences}
\end{figure}
\begin{figure}[!b]
	\includegraphics[width=\linewidth]{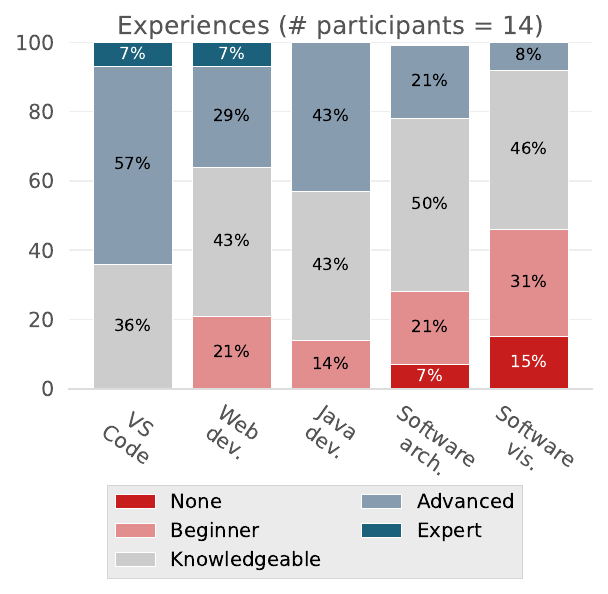}
	\caption{Participants’ reported experiences for different aspects.}
	\label{fig:pre-experiences}
\end{figure}
\subsection{Participants}
We invited students of Kiel University that attend the Bachelor's or Master's program in computer science to participate in our user study~\cite{falessi2018UseStudents}.
The participation was voluntary.
All participants could sign up for random group assignment or participate with a fellow student.
Each group had the chance to win two out of ten 100 € gift cards for an e-commerce shop~\cite{camerer1999Incentives}.

\paragraph*{Distribution}
The conducted user study included seven groups with two students each.
The number of participants is therefore slightly larger than the median participant count (thirteen) in the related literature body~\cite{merino2018SLR}, but too small to be effectively used in a controlled experiment~\cite{gall1996, cohen2017}.
With the exception of one group, all other participants within their group knew each other.
Five students attend the Master's program, the remaining students are undergraduates in computer science.
All participants reported that they intend to become professional software developers. 

\paragraph*{Experiences}
Figure~\ref{fig:se-experiences} shows participants' reported experiences with software development based on work experiences.
The two students who indicated they had no experience are in the undergraduate program, and one of them also indicated (as only person) that the decision to become a software engineer is not final.
The remaining twelve participants have either gained experiences while working as student employee or in private software development.
Three participants are additionally involved in open source development.
Figure~\ref{fig:pre-experiences} shows the results of various experiment-related aspects that were asked.
All participants stated that they have knowledgeable or even better experiences in VS Code.
Three persons rate their web development and software architecture experiences at beginner level.
One of the participants with no software engineering work experience reported to have no experience in software architecture.
Overall, the distribution of experiences match the courses of study, since SVs are often treated as seminar papers in the master's program, for example.
However, we probably also see overestimation such as the persons that stated to be at expert level for VS Code and web development, as well as half of the participants stating to have at least knowledgeable experiences in SV.
In this context, half of the participants have used ExplorViz at least once in the past.
The participants of three groups each have different experiences with ExplorViz.

\subsection{Target System and Task}\label{subsec:targetsystem}
ExplorViz' SV visualizes a software system's runtime behavior.
However, it is not limited to application tracing of monolithic software systems, but also supports distributed tracing,\footnote{\url{https://opentelemetry.io}} e.g., network requests between applications that use distributed architectures.
Since distributed software systems are pretty common nowadays, we incorporated this fact in our experiment.
To achieve that, we used the distributed version of the Spring PetClinic\footnote{\url{https://github.com/spring-petclinic/spring-petclinic-microservices}} as target system for the experiment.
As done in the past~\cite{krauseglau2022vissoft} we recorded traces during the execution of use cases within the PetClinic.
For the experiment, these were then provided as so called snapshots, i.e., aggregated runtime behavior, to the \textsf{Frontend},  resulting in a structural, `static' SV of dynamic runtime behavior.
We decided against using multiple snapshots so as not to overwhelm new users with the amount of features.
However, this can be seen in the supplementary video of this work.
The participants explored the target system by means of its source code as well as embedded SV and were asked to solve two tasks.
\begin{table}[h]
	\centering
	\caption{Program comprehension tasks that participants had to solve..}
	\begin{tabular}{l l p{0.24\textwidth}} \toprule
		ID & Category & Question \\ \midrule
		T1  & Structural Understanding & What do you think is the reason that the `Owner' class is instantiated multiple times, but the other classes in the relevant program flow are instantiated only once? \\\\
		T2  & Software Insight  & Name all Java classes that are involved in a program flow to show the visit screen with the new `select veterinarian' feature.  \\\bottomrule
	\end{tabular}	
	\label{table:tasks}
\end{table}

Table~\ref{table:tasks} depicts the program comprehension tasks that all participants had to solve during the experiment.
We did not use metric analysis tasks such as `find the class with the highest instance count'.
Instead, the chosen tasks instructed the participants to structurally comprehend the software and find analogies based on the depicted runtime behavior.
Therefore, the tasks of the experiment refer to a scenario as presented in SC1, i.e., a guided, task-oriented introduction for onboarding.
With the focus on SC1, we intend to investigate both the non-collaborative and collaborative onboarding process.
Therefore, T1 had to be solved alone and served as an introduction to both the target system and the use of our approach.
T2 introduced the collaborative features, e.g., shared SV and synchronized text selection events, and asked the participants to work together.

\subsection{Procedure}
In the following, we present the experiment's procedure.
For additional information, we refer readers to the second prepared video\footnote{\url{https://youtu.be/wdkeDDPXeQQ}} that demonstrates an exemplary experiment run.
Overall, the experiment is divided into pre-questionnaire, mid-questionnaires, i.e., questions that had to be solved after each task completion, and post-questionnaire.

The user study took place at Kiel University and included one instructor who also co-authored this paper.
The instructor designed and implemented the approach as well as conducted the user study.
Although our approach can be used remotely, we decided to have the study take place in one locality, so that the instructor could intervene if necessary.
In each experimental run, the participants were first informed about the data that would be recorded and used for publication.
After signing a consent form, the instructor gave a brief introduction to VS Code and the embedded SV.
It was mentioned that all introduced features were additionally described on a cheat sheet, which was placed on the table in front of the subjects. 
Afterwards, the participants were told to openly ask questions if they had a problem.
Furthermore, they were told that they could pause or abort the experiment at any time.
They draw their login token for the survey tool LimeSurvey\footnote{\url{https://www.limesurvey.org}} and started with the pre-questionnaire.
Then T1 was introduced and all participants were redirected to browser-based VS Code instances by clicking a button inside of the LimeSurvey form.
Each VS Code instance was specifically prepared for a given task and ready to use.
It did not require any setup, so that the participants could completely focus on the task itself.
They began by reading a markdown file that introduced the target system, controls, and the task itself.
After answering T1 in LimeSurvey, all participants gave their feedback to the just used approach. T2 was introduced in the same way as T1.
However, here participants were instructed to test the collaborative features first and then work together on solving T2.
Again, the subjects gave their feedback and concluded with the post-questionnaire.
During each experiment run, the instructor made a note of noticeable mentions stated by the participants.
	\section{Results \& Discussion}\label{sec:results}
Our mid-questionnaires and post-questionnaire contained statements for which participants had to indicate their level of (dis)agreement on a 5-point Likert scale.
The questionnaires also included free reply fields to leave a comment on any experiment-related matter.
Additionally, the instructor made a note of observations such as rational usages of specific features as well as noticeable emotions~\cite{merino2018SLR} and mentions of the participants.
In the following, we present and use the results of our conducted user study to revisit our posed research questions.
Furthermore, we discuss the threats to validity of our evaluation.
Although we use the term SV in this paper, we do not want it to be understood as a generalization of our results.
We again emphasize that the results and their interpretation are restricted to our particular prototype using collaborative code cities and our experiment.
Therefore, the findings should be seen as first insights and indicators for refinements rather than statistically grounded results.

\begin{figure}[b]
	\includegraphics[width=\linewidth]{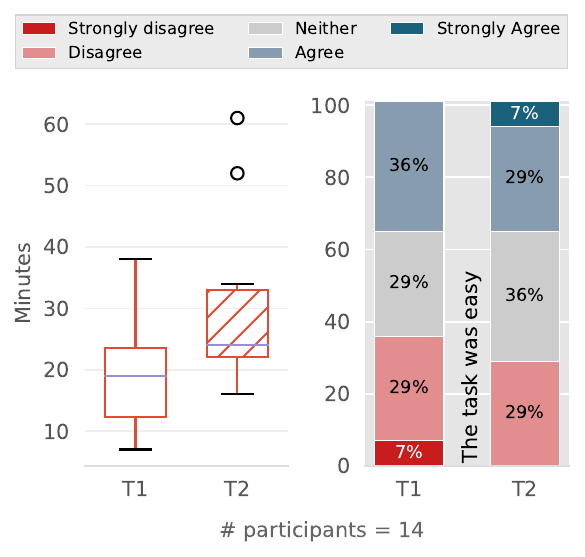}
	\caption{Total time spent \& perceived difficulty per task.}
	\label{fig:time}
\end{figure}

\subsection*{Task evaluation}

We measured an overall task correctness of 90 \%.
The related time spent solving the tasks is depicted in Figure~\ref{fig:time}.
The average time spent on T1 is for both the mean and median 19 minutes.
The fasted participant correctly solved T1 in seven minutes.
This person was already familiar with ExplorViz.
For T2, we see 29 minutes for the mean and 24 minutes for the median.
Both tasks were without time limit, hence the outlier group for T2.
Figure~\ref{fig:time} also depicts the participants' perceived task difficulty.
T1 and T2 were found to be difficult by four participants, with T1 also found to be very difficult by one person.
Due to the overall distribution, we conclude that the tasks were neither too easy nor too difficult.

\subsection*{\textbf{RQ1}: How do subjects use the embedded SV and code editor during task solving?}
\begin{figure}[t]
	\includegraphics[width=\linewidth]{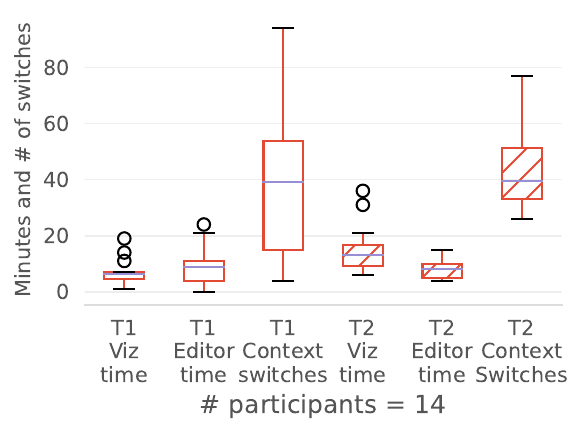}
	\caption{Time spent per tool \& number of context switches performed per task.}
	\label{fig:context-switches}
\end{figure}

To the best of our knowledge, this work presents a novel approach that combines code editors with remote pair programming techniques and embedded, collaborative code cities.
Therefore, we first intend to understand how the participants in our study use the approach with free choice of the tool, i.e., embedded SV and code editor, as well as with tasks referenced to SC1.
In that context, Figure~\ref{fig:context-switches} depicts the time spent using each tool per task.
For measurement, a VS code event was used to capture the time at which participants clicked on the code editor or the ExplorViz extension, therefore switched their focused context.
We would like to mention that it was technically (due to VS Code's limitations for extensions) only possible to measure the time spent between context switches.
Thus, if a participant did not change the context but, for example, only used the SV, then our measurements indicate a time spent of one minute for the SV.
This is the case for the fastest participant for T1 mentioned above, who actively interacted only with the SV during this task (as confirmed by the video recording).
The average time spent using the SV for T1 is seven minutes and nine minutes for VS Code (both mean and median).
During this task, participants comprehended the source code for the first time and probably spent more time reading it.
It is therefore surprising that the time difference for the first task is already quite small. 
The reason for this is that code cities can facilitate the understanding of structures and are therefore suitable for the task of obtaining an overview~\cite{wettel2011ControlledExperiment, galperin2022}.
This was also explicitly mentioned by three participants in the free text fields.
For T2,  the average time spent using the SV is fifteen minutes and eight minutes for VS Code.
The (almost) double amount of time spent using the SV results from the two outliers.
For this task, however, the median for time spent using the SV is thirteen minutes and eight minutes for VS code.
We suppose that this comes from the shared software cities and the ability to highlight objects in question.
The instructor's notes mention the frequent use of shared popups within two groups.
The video recordings confirm that these groups often use the popups as a basis for discussion.
Also, participants often use the ping feature of our tool to highlight certain details for their collaborator.
Therefore, they spent more time using the SV.
However, collaboration is not the only reason for that.
T2 explicitly requires to understand and extend a program flow.
The SV provides a visual overview of the software system's structure and in our case also of a runtime behavior snapshot (see Section~\ref{subsec:targetsystem}).
As a result, it is far easier and obvious to use this available visualization and for example trace imaginary method calls with the mouse cursor
(especially, when combined with collaborative features).

Figure~\ref{fig:context-switches} also presents the number of context switches for each task.
We observe that for T1 the number of switches between SV and code editor is much more distributed among the participants than for T2.
Again, the reason for that is presumably the collaboration in T2.
Most of the time, the participants work together and therefore change their tool when initiated by the other collaborator.
For both T1 and T2, the median of context switches is around forty, indicating that the amount of context switches is independent on our tasks and collaboration.

Since our approach incorporates the runtime behavior of the target system, we also intended to know how participants perceived the usefulness of the two tools to comprehend the posed program flow of T1.
In this context, Figure~\ref{fig:t1} shows that the SV was perceived as more useful than the code editor.
One participant mentioned that the communication lines are one of the most beneficial properties of the SV.
In ExplorViz, the communication lines incorporate runtime information such as the method call's frequency in the visualized snapshot.
These information are important to comprehend runtime behavior.
Additionally, the SV already maps the runtime information that the users would otherwise have to find and understand on their own.

\begin{tcolorbox}
	We conclude that the participants used the SV as supplement to the code editor for specific comprehension tasks.
\end{tcolorbox} 

\subsection*{\textbf{RQ2}: Is the code editor perceived as more useful than the embedded SV?}
Traditionally, understanding a software system's behavior is primarily achieved by comprehending the source code~\cite{xia2018}.
For this experiment, the results related to RQ1 show that our approach was, for example, used by the participants to gain an overview of the target system.
This is a common and suitable use case for SV, as shown in the past~\cite{wettel2011ControlledExperiment}.
However, professional developers question the need for SV~\cite{siegmund2016,sensalire2008}.
In our opinion, one of the reasons for that is the lack of properties such as code proximity~\cite{bassil2001, kienle2007} and the SV tool's setup~\cite{sensalire2007}.
In that context, we now examine how participants rate the usefulness of our approach.

\begin{figure}[t]
	\includegraphics[width=\linewidth]{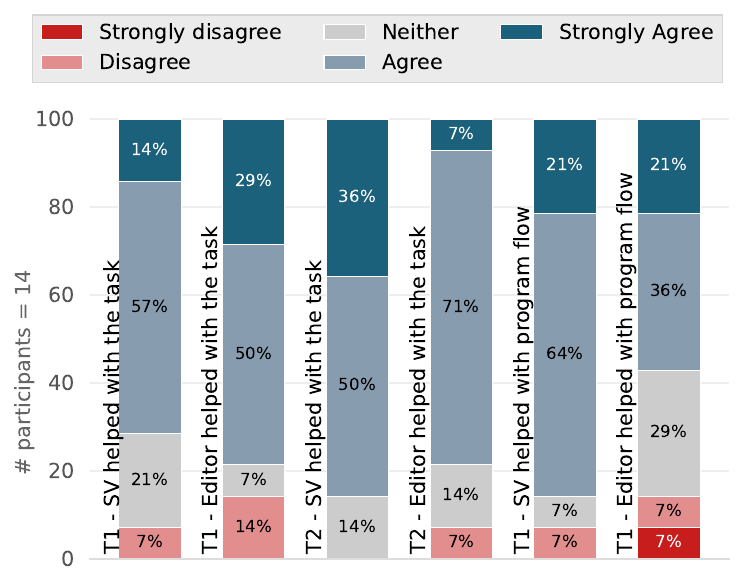}
	\caption{Mid-questionnaires - Perceived usefulness for tasks}
	\label{fig:t1}
\end{figure}

Figure~\ref{fig:t1} depicts the results of the mid-questionnaires regarding the perceived usefulness of the tools for a task. 
For T1, overall 71 \% agree with the posed statement `SV helped with the task'.
The usefulness of the code editor was slightly (one person difference) more agreed to.
However, for the SV the number of participants who neither agree nor disagree is higher and those who disagree is lower.
Regarding T2, we see that overall 86 \% agree with the posed statement `SV helped with the task'.
In comparison,  the code editor's usefulness was slightly (one person difference) less agreed to.
\begin{tcolorbox}
	We conclude that the participants perceive code editor and SV as approximately equally useful (in the context of the task solving).
\end{tcolorbox} 

\subsection*{\textbf{RQ3}: Do subjects recognize the usefulness of collaborative SV features for specific tasks?}

With RQ3, we expand the results of our previous work~\cite{krauseglau2022vissoft} regarding the perceived usefulness of collaborative code cities.
In this context, we asked all participants to state their level of agreement with two statements posed.

\begin{figure}[t]
	\includegraphics[width=\linewidth]{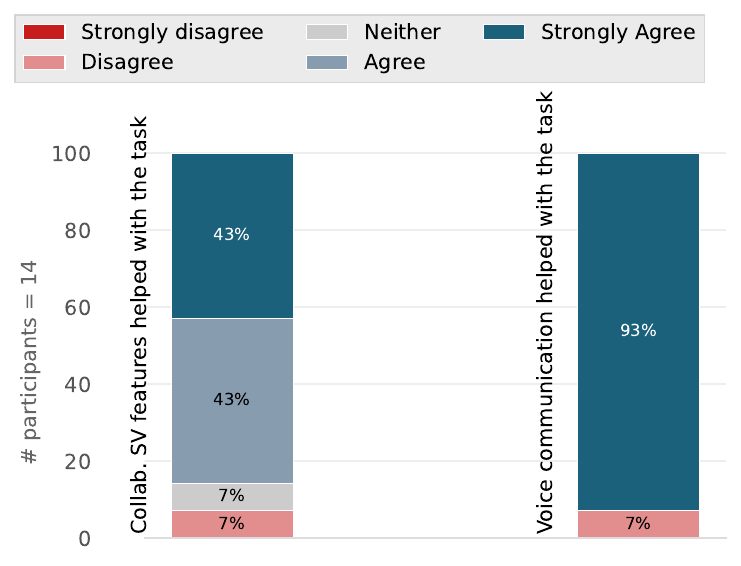}
	\caption{Mid-questionnaire - T2 - Collaboration}
	\label{fig:t2}
\end{figure}

Figure~\ref{fig:t2} presents the related results.
We see that 43 \% of the participants agree or strongly agree with the statement `Collaborative SV features helped  with the task', respectively.
The one person that disagrees with the statement mentioned that the collaborative SV features did not help in his case, since there was barely any input from the other participant. 
However, he agrees that the communication would be a big help in pair programming supported development in the real world.
Presumably due to the low contribution of his collaborator, the same person also disagrees with the second statement that is concerned about voice communication.
Due to low input from his collaborator, the same person also disagrees with the second statement, which refers to the perceived usefulness of voice communication.
Nevertheless, all of the remaining thirteen participants strongly agree that voice communication was helpful in the task.
This is consistent with our previous findings indicating that voice communication is one of the most useful collaborative tools in SV~\cite{krauseglau2022vissoft} .
\begin{tcolorbox}
	We conclude that the majority of participants perceive the collaborative SV features as useful in the given task.
\end{tcolorbox} 

\begin{figure}[h]
	\includegraphics[width=\linewidth]{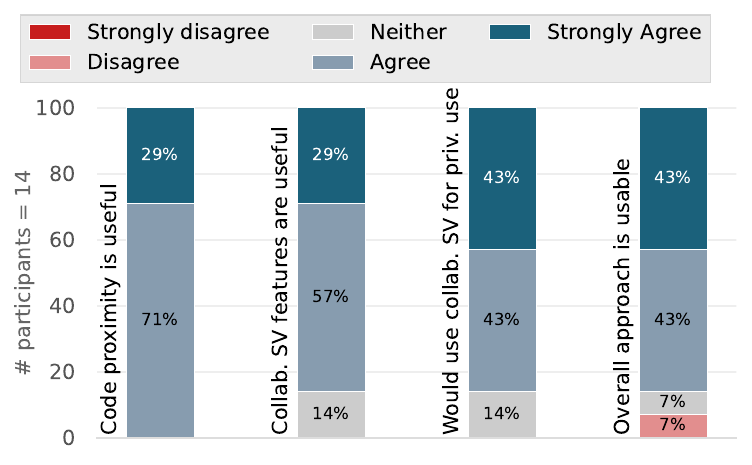}
	\caption{Post-questionnaire - Perceived usefulness and usability}
	\label{fig:post}
\end{figure}

\subsection*{\textbf{RQ4}:  What is the general perception of the usefulness and usability of the approach?}
The post-questionnaire was designed to capture participants' overall perceptions of the approach's usefulness and usability.
By answering RQ1, we have seen that the participants indeed use the SV as supplement during the comprehension task.
For RQ2, we concluded that participants perceived code editor and SV to be about equally useful in the context of a real-world task.
Finally, Figure~\ref{fig:post} shows:
\begin{tcolorbox}
	All participants agree or strongly agree that the SV's code proximity is generally useful.
\end{tcolorbox} 
Collaboration is obviously dependent on many factors, e.g., mutual perception of collaborators or motivation.
In our context, we have seen this for RQ3 or in previously published results~\cite{krauseglau2022vissoft}.
The participants rate the collaborative SV features slightly different when to be evaluated independently of a task.
Figure~\ref{fig:post} shows a shift in the distribution of approval ratings.
The one person who previously disagreed with the usefulness of the collaborative features now neither agrees nor disagrees.
That fits his previous mentions.
Compared to perceived usefulness for T2, overall perceived usefulness of collaborative SV features shows less strong agreement.
As a matter of fact, we could not find a reason why two participants downgraded their level of agreement to `agree'.
However, the overall approval rate remains the same.
\begin{tcolorbox}
We conclude that the majority of subjects perceive the collaborative SV features as useful.
\end{tcolorbox}

Although this evaluation is overall more concerned about the perceived usefulness of embedded SV, identified usability problems can help to identify desirable refinements.
In this context, Figure~\ref{fig:post} also presents the participant's perceived usability of our approach.
The results show that 86 \% of the participants find the used combination of embedded SV and code editor usable.
There are some desirable improvements that are mentioned via text response, e.g., better performance.
However, the biggest usability problem was the unintended minimization of the embedded SV.
The reason for that is that VS code opens files that have been clicked in the package explorer in the currently focused editor group.
This behavior can be disabled by locking an editor group.
However, at the current time of writing, the lock mechanism cannot be triggered from within a VS Code extension.
Figure~\ref{fig:post}  also shows that another 86 \% would use this approach for private purposes such as collaborative program comprehension with fellow students.
\begin{tcolorbox}
	We conclude that the majority of participants find our approach usable.
\end{tcolorbox}

\begin{figure}[t]
	\includegraphics[width=\linewidth]{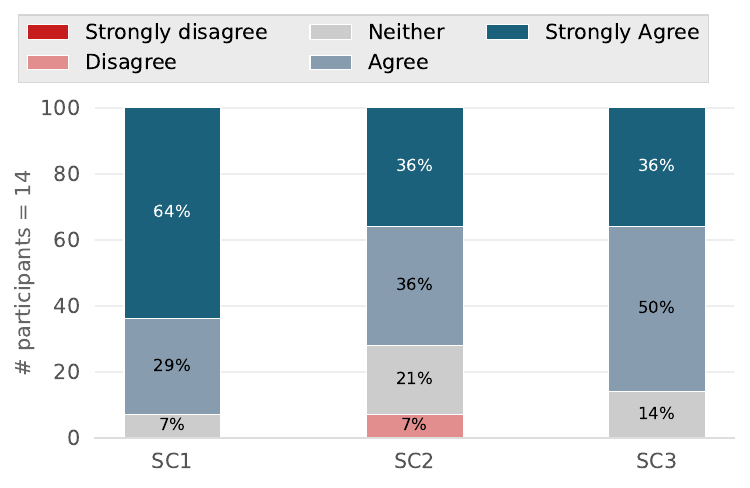}
	\caption{Post-questionnaire -- Perceived usefulness of the approach when applied in a described scenario. (see Section~\ref{sec:scenarios})}
	\label{fig:post-scenarios}
\end{figure}

\subsection*{\textbf{RQ5}: Is the approach perceived as useful in the envisioned usage scenarios?}
Our pilot study found that a single experiment run would take about an hour to complete.
In order not to discourage potential participants due to the time to be spent, we decided to ignore the other usage scenarios and only use tasks in the experiment based on SC1.
Nevertheless, the post-questionnaire was also used to capture participants' perceived usefulness in applying the approach in the remaining, envisioned scenarios.
In this case, they were described in text and subjects were asked to state their agreement on a 5-point Likert scale.
Figure~\ref{fig:post-scenarios} depicts the related results.
The complete scenario descriptions are available in the supplementary package of this paper~\cite{krause2023CollabSuppl}, but essentially summarize the envisioned usage scenarios in Section~\ref{sec:scenarios}.
The participants rated SC1 with the highest overall agreement and strong agreement, respectively.
The experiment's tasks and their introduction originate from SC1.
SC2 has the highest amount of neutrality and disagreement.
One person that answered with neither agreement nor disagreement mentioned that code changes are usually reviewed before deploying them.
Since our approach only shows runtime behavior, he is not sure how changes will be visualized for the code review.
This detail was in fact omitted in the textual description of SC2.
We believe that this uncertainty is the reason for the highest amount of neutrality and disagreement for SC2.
However, the majority consensus was positive for all scenarios.
\begin{tcolorbox}
	We conclude that the majority of subjects find the application of our approach useful in the posed scenarios.
\end{tcolorbox}

\subsection{Threats to Validity}


\paragraph*{Remote pair programming solution}
As mentioned in Section~\ref{sec:implementation}, we decided to implement our own remote pair programming approach, so that the reproducibility of our evaluation is not dependent on the availability of external services.
However, this custom implementation lacks useful features compared to full-fledged solutions for remote pair programming.
For example, one participant mentioned that he was unable to draw the attention of the collaborator to a specific code part.
Although our study did not aim to evaluate whether one tool is better than the other, this custom implementation may have influenced the perceived usefulness or usability of the SV or code editor.
In contrast, Figure~\ref{fig:t1} shows that the participants find the SV to be more suitable to understand dynamic program flows.
With that being said, we conclude that more empirical research is required in this context.

\paragraph*{Experiment duration}
The average time spent on the user study was about one hour, both median and mean.
It follows that the attention span of the participants and thus the results might have been influenced.
To mitigate this, we told participants during the introduction that breaks could be taken at any time and participation could be aborted. 
Moreover, T2 was solved collaboratively and therefore presumably relieved the experimental situation.  

\paragraph*{Target system}
The prepared target system contains 26 application logic-related Java files that are distributed among four Maven subprojects.
As a result, the small project size may have influenced the perceived usability of the SV, as also mentioned by one participant.
We agree, but also emphasize that we did not intend to evaluate usability based on the scalability of the visualization, but on the overall concept.
Overall, this evaluation is more concerned about the perceived usefulness of SV incorporating distributed tracing for the onboarding process.
In addition, we argue that a real-world application of the onboarding scenario with SV should guide new developers through the software system's behavior with increasingly large portions of the code base.

\paragraph*{Participants}
The use of students in experiments is a valid simplification that is often said to possibly compromise external validity~\cite{falessi2018UseStudents}.
In our case, the participants' experiences might have influenced their perception regarding the usefulness of the SV as well as their time spent using the SV.
In this context, professional developers can benefit from their experience, e.g. with the Spring framework, and can understand the source code faster.
As a result, we will repeat the experiment with professional developers.


	\section{Related Work}\label{sec:related-work}
Code proximity is often not clearly emphasized in SV publications, but follows from the mentions of code viewers in the text itself.
Therefore, there are numerous research approaches that use embedded code viewers within SV such as code cities~\cite{wettel2011ControlledExperiment}.
This also applies to more recent and often collaboratively usable virtual reality approaches~\cite{koschke2021SEE,Oberhauser_Lecon_2017,khaloo2017,hori2019CodeHouse,merino2017CityVR}.
Other publications present different use cases for embedded SV in code editors, such as dependency browsing~\cite{borowski2022GraphBuddy} or sketching~\cite{lichtschlag2014,kurtz2011CodeGestalt}.
Few approaches enable developers to modify source code via embedded code editors~\cite{dominic2020PairProgrammingVR}.
Due to space limitations, we cannot address and discuss all related approaches~\cite{lintern2003,sulir2018,kurbatova2021}, but focus below on what we consider to be the most comparable work.

In 2015, Balogh et. al presented a refined version of their tool CodeMetropolis~\cite{balogh2015codemetropolis}.
Their approach uses static analysis of a software systems' source code and visualizes the result as 3D code city.
The related rendering is achieved using a modded version of the video game Minecraft.
Thanks to the multiplayer mode of Minecraft, the code city can also be explored collaboratively.
Overall, CodeMetropolis and ExplorViz share the aspects of collaboration and code editor integration.
However, these features are implemented differently in each case.
For example, in CodeMetropolis users navigate through the same instance of a given code city using the first-person perspective.
They can see the avatars of collaborators and interact based on Minecraft's limitations.
In ExplorViz, the collaboration is achieved using collaborative SV features, e.g., shared popups.
Regarding the code editor integration, both CodeMetropolis and ExplorViz provide an extension than can be installed in Eclipse and VS Code, respectively.
In this context, both extensions provide a comparable set of features, e.g., open Java class in the SV.
However, our extension is also able to embed the SV in the actual code editor, whereas the Metropolis approach can only be used as external SV that links to the code editor (see Section~\ref{sec:design}).
	\section{Conclusions \& Future Work}\label{sec:conclusions}
In this paper, we presented the architectural design of our approach for collaborative, code-proximal dynamic software cities within code editors.
The main idea is to link collaborative SVs directly to the source code that is under development within a code editor and vice versa.
We have prototyped this approach within our SV tool ExplorViz.
The result is a VS Code extension that either embeds three-dimensional software cities in the code editor or acts as gateway between code editor and external SV.
Therefore, we directly link runtime behavior with the related program elements, for example, Java methods.
Users can collaboratively explore the SV from within their code editor using synchronized software cities and collaborative SV features, e.g., shared popups.
In addition to the implementation, we sketched three envisioned usage scenarios.

We conducted an initial user study to collect first-time feedback regarding the perceived usefulness and perceived usability of our approach.
The results show that the majority of participants generally perceive the approach as useful and usable.
In this context, participants rated code editor and SV as equally useful in solving the given program comprehension tasks.
The measured time spent in each tool, i.e., SV and code editor, indicates that the participants indeed use the SV as supplementary tool.

In the future, we will implement useful features and refinements.
Additionally, we plan to repeat the experiment with professional developers.
	
   \section*{Acknowledgment}
   The authors would like to thank Malte Hansen and Lennart Ideler for their contributions with implementing and evaluating some of the features presented in this paper.
	
	\clearpage
	\providecommand{\doi}[1]{DOI: \href{https://doi.org/#1}{#1}}
	\bibliographystyle{myIEEEtran}
	\bibliography{jabrefdb-akr}
	
\end{document}